\documentclass[conference]{IEEEtran}

\usepackage{multirow}
\usepackage{booktabs}
\usepackage{color, colortbl}
\usepackage{placeins}
\usepackage{graphicx}
\usepackage[hidelinks]{hyperref}
\usepackage{xurl}
\usepackage{cite}
\usepackage{balance}
\usepackage{threeparttable}
\usepackage{nicematrix}
\usepackage{placeins}
\usepackage{algorithm}
\usepackage[noend]{algpseudocode}

\begin{document}

\title{Work in progress: Identifying Two-Factor Authentication Support in Banking Sites}

\author{\IEEEauthorblockN{Srivathsan G. Morkonda$^*$ and AbdelRahman Abdou$^\dagger$}
\IEEEauthorblockA{\textit{School of Computer Science} \\
\textit{Carleton University, Ottawa, Canada} \\
\textit{Email: $^*$srivathsan.morkonda@carleton.ca, $^\dagger$abdou@scs.carleton.ca}}}

\maketitle

\begin{abstract}
Two-factor authentication (2FA) offers several security benefits that security-conscious users might expect from high-value services such as online banks. In this work, we present our preliminary study to develop a scoring scheme to automatically recognize when bank sites mention support for two-factor authentication. We extract information related to security features (primarily 2FA) offered by 379 bank domains from 93 countries. We use a subset of these sites to refine our scoring scheme to include several heuristics for identifying whether sites offer 2FA. For each bank domain in our dataset, we use our algorithm based on text-analysis to calculate whether the domain offers 2FA to the users of the domain's online banking platform. Our preliminary findings suggest that 2FA is yet to be widely adopted by banking domains.
\end{abstract}

\section{Introduction}
Users of online services are often encouraged to use two-factor authentication (2FA) for additional security of their accounts. In addition to what the user knows (e.g., a password), requiring the user to provide a second factor of a different defence dimension~\cite{alaca2019comparative}, or something the user has (e.g., a hardware token), can increase security as it prevents attackers from exploiting guessed/leaked passwords~\cite{van2020computer}. This additional protection is especially important in high value services such as financial user accounts. With over 1.9 billion users (expected to grow to 2.5 billion by 2024)~\cite{statistaMarket} of online banking services, 2FA is an important security requirement to protect financial accounts. Users could consider several factors when choosing a bank that offers online services. While usability is often an important consideration, security-conscious users might expect the banking platform to offer 2FA. However, not all online banking service providers offer 2FA to their users.

Banking services often provide information about their web platform's security features within their sites. For example, sites might use their frequently asked questions (FAQ) page to explain to users about various security features, including 2FA (e.g., how users can setup 2FA for the site). Such pages can also be useful for extracting information about a site's support for 2FA (if any) and the type of 2FA offered, such as SMS one-time password (OTP) or a security device. In this work, we aim to develop a measurement methodology to automatically identify whether a site offers 2FA. We are interested in consumer banking sites as users of such services might expect 2FA security features. Our contributions include:
\begin{itemize}
    \item A measurement methodology to scan web pages to identify mentions of 2FA support for online bank accounts.
    \item An empirical study to estimate the number of online banking sites around the world that offer 2FA for online bank accounts.
    \item Additionally, a custom crawler optimized for crawling security-related pages that might contain information about various security features offered by a site.
\end{itemize}

In the next section, we discuss our methodology for building our dataset on bank domains around the world, and explain our scoring algorithm for rating sites on 2FA support. In Sec.~\ref{sec.empirical.study}, we report the findings on the number of bank sites that offer 2FA based on our scoring algorithm. Related work is discussed in Sec.~\ref{sec.related.work}. Sec.~\ref{sec.future.work} explains our next steps in the project. Our preliminary conclusions are included in Sec.~\ref{sec.conclusion}.

\begin{figure*}[tb]
    \centerline{\includegraphics[width=0.9\textwidth]{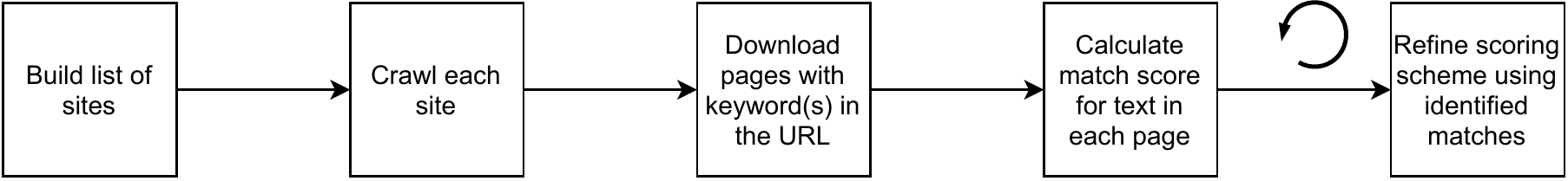}}
    \caption{Overview of our methodology to identify mentions of 2FA feature offered in a site}
    \label{figMethodologyOverview}
\end{figure*}

\section{Methodology}
In this section, we describe the steps we took in our study to analyze 2FA support in online banking services. Our workflow is shown in Fig.~\ref{figMethodologyOverview}. First, we automatically collect a list of links to online banks in various regions of the world (Sec.\ref{sec:initiallist}). Then, using a custom-built crawler, we searched each bank site to download pages with mentions to the site's security features~(Sec.\ref{sec.extracting.pages}). Using this dataset, we iteratively developed a scoring algorithm to identify whether a site offers 2FA~(Sec.\ref{sec:2famention}).

\subsection{Our measurement dataset}
\label{sec:initiallist}
\begin{figure}
    \includegraphics[width=\columnwidth]{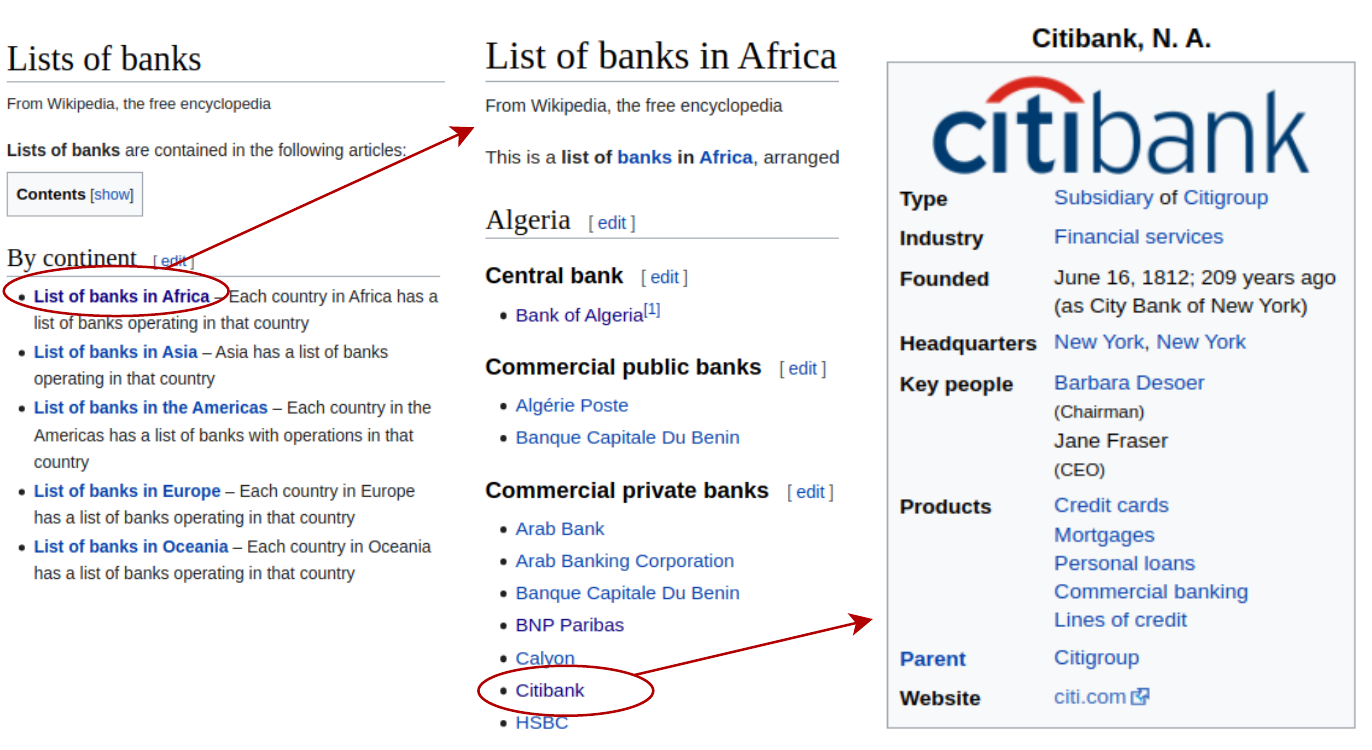}
    \caption{Steps for extracting the bank name and site link for an example bank site. For each region (left), we follow the links listed under each country (middle) to visit the individual bank pages (right).}
    \label{figCitiPage}
\end{figure}

For our measurement, we gathered a list of domains for 751 bank sites around the world. As shown in Fig.~\ref{figCitiPage}, starting from the Wikipedia page of a list of bank sites across various countries~\cite{wikiBankList}, we automatically followed each listed link to the individual page of the bank. Within each page that provides information about the bank, we extracted the bank name and a link to the bank site. This list also includes links to central banks that typically do not offer online accounts for individual users. For each entry of a central bank, there are several consumer banks (the focus of our study) listed under the country. Although not every consumer bank might be included, the collected links provide a representative list of consumer banks from around the world. If a site is listed under multiple countries (i.e., in each country where the bank has a presence), we only consider it once and append a label for each country it is listed under.

Our goal is to extract as many relevant pages as we can to later scan the content within each page for mentions of 2FA support. Our approach for identifying 2FA support relies on language semantics used by banking platforms. In this study, we limit our data collection and analysis to content in English. If a site supports multiple languages including English, our data collection tool can obtain the relevant content if a link to the English version exists in the site's landing page.

\subsection{Finding security-relevant pages}
\label{sec.extracting.pages}
We developed a tool built on top of Scrapy~\cite{scrapy} to recursively crawl the web pages of a bank website. We crawl pages that likely contain information related to the site's security features. We only follow links that belong to the website being crawled. For example, if we are crawling \texttt{bankofamerica.com}, we follow links like \texttt{about.bankofamerica.com} and \texttt{bankofamerica.com/credit-cards}, but not \texttt{example.com}. For our initial study, we collected data from 751 bank domains to obtain the HTML content for each page identified as relevant. The following design choices helped reduce the crawl time:

\begin{itemize}
    \item As we limit our study to only pages in English, we use the texts in buttons and hyperlinks of a page to empirically determine whether the page is mainly in English. In our initial analysis, we find that non-English pages tend to have fewer than 25\% of texts in English. For example, some common English words found in non-English pages include ``Cookies", ``Login", and ``Contact". For each page, we collect the list of words (splitting texts with hypen (-), colon (:), and space) excluding special characters and numbers to check if more than 50\% of the words are found in a common dictionary of English words. For pages that have fewer words, we stop crawling beyond the page and discard the page as non-English.
    
    \item We skip URLs that are likely to be unrelated to the site's security features. For each page, we check the URL for keywords that suggest that the page provides information unrelated to our study such as details about the bank's leaders (executives). In an initial analysis, we identified crawls that include large groups of URLs that are clearly not related to 2FA. Such unrelated groups lead to many unnecessary scans that slow down the crawl. Based on our analysis, we terminated a crawl branch when the URL contained any of the following strings: \texttt{"news", "media", "people", "team", "leaders", "career", "forum", ".pdf", and "tel:+"}.
    
    \item We skip previously crawled links to prevent scanning pages repeatedly. For example, bank sites often include helpful links such as ``Contact us" in multiple pages to help users navigate easier. Without checking for revisited links, our tool might unnecessarily crawl common links from each page.
    
    \item We limit the crawls to a maximum depth of 10, starting from the landing page in order to prevent crawls leading into a long chain of links. We find that some crawls tend to get into lengthy links (e.g., to bank reports with date structures) that stall the crawling program. If a site makes no mention of 2FA within a depth of 10 from the landing page, we assume that users might not see this information as an available feature.
\end{itemize}

To avoid placing undue load on domains, we limit the number of simultaneous requests on each domain to seven. To collect data more quickly, we ran multiple instances of our tool to process separate lists of domains simultaneously. It is likely possible to improve further by identifying other crawl groups to exclude. But our current improvements proved sufficient for collecting data within reasonable runtimes.

\begin{figure*}[tbh]
    \includegraphics[width=2\columnwidth]{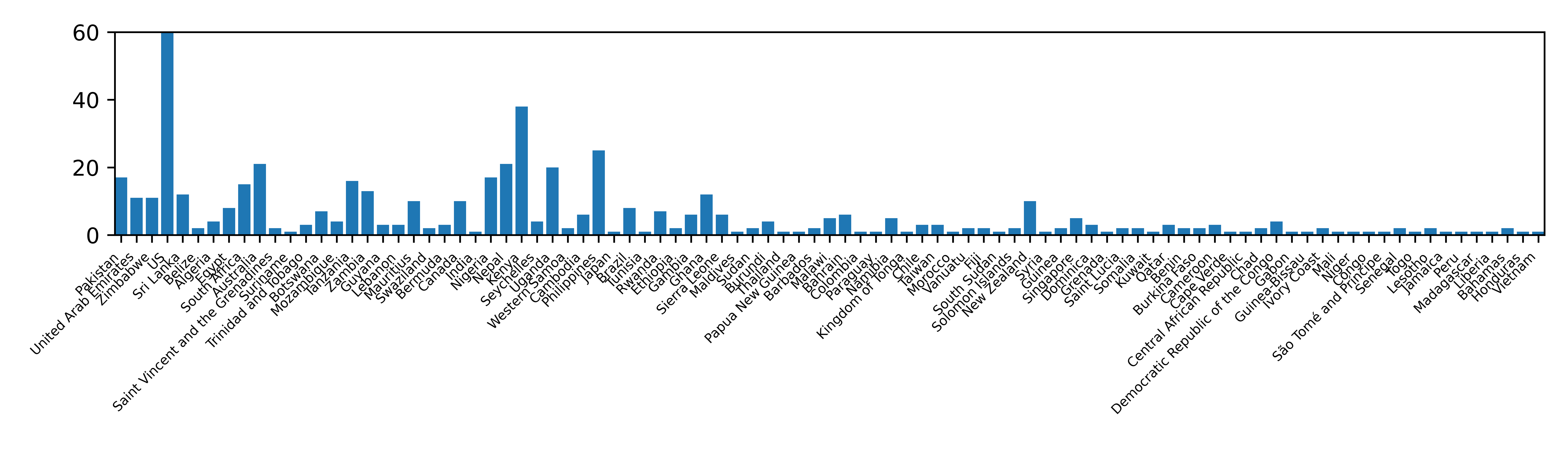}
    \caption{The number of banks in each country. In total, we analyzed 2FA support on 379 bank websites from 93 countries.}
    \label{figCountries}
\end{figure*}

\subsubsection*{\textbf{Matching URL}}
We identified keywords commonly found in the URLs of pages that mention two-factor authentication support. Our list of the keywords include \texttt{"security", "privacy", "two-factor", "faq", "authentication"} and other variations of these words (e.g., \texttt{"two-step"} and \texttt{"2fa"}). For URLs that contain these keywords, we saved the page locally to be included for analysis of the bank domain. As a typical crawl example, for \texttt{chase.com} we crawled 19,244 links in 18.5 minutes and downloaded 195 URL match web pages---i.e., all of which contained one of the URL keywords above, including pages mentioning the platform's support for 2FA.

\subsubsection*{\textbf{Dataset for text analysis}}
We did not find any English pages for 226 of the 751 banks. Among the remaining 525 sites with one or more English pages, we did not find a URL match for 146 (or~$\sim28\%$) sites. 
For these domains, we cannot determine the security features they offer. It may be that these sites use other URLs for providing information about security.  While it may be possible to download and process all pages from each domain, it complicates the analysis by including non-security pages that might use terms similar to ``two-factor authentication" (e.g., information about ``multi-factor investing").

We found URL match web pages for 379 banks in 93 countries (see Fig.~\ref{figCountries}). The number of such pages for each bank varied from one (56 banks) to 3,143 (Century Commercial Bank in Nepal). The total number of these pages was 16,077, occupying 3.5GB (JSON formatted).

\subsection{Identifying mentions of 2FA}
\label{sec:2famention}
By looking for text patterns with mentions to specific security features, we can guess whether the site offers that feature. For example, if a page includes the text ``Our online platform supports two-step verification", we can guess with higher probability that 2FA is offered on the site. 

We develop a scoring algorithm for rating 2FA support, and empirically pick the threshold through which we aim to determine whether a bank supports 2FA. We consciously decided to begin our journey in this project by pursing such a heuristics-based path as one that we can well understand. We plan to explore other approaches, including, e.g., ML-based classification, in the future. 

\begin{algorithm}[hbt!]
\caption{Calculating match score for a website}\label{alg:page.score}
\begin{algorithmic}
\State $max\_score = 0$
\For{\texttt{$i$ in URL match web pages}}
	\If{$url(i)$ in \textit{IGNORE\_LIST}}
	    \State \textbf{skip} $i$
	\EndIf
	\State $score = 0$
	\If{$url(i)$ matches \textit{URL\_REGEX}}
	    \State $score = 0.25$
	\EndIf
	\State $sentences = \texttt{pre\_process}(raw\_html(i))$	
	\For{\texttt{$j$ in $sentences$}}
	    \If{$j$ in \textit{NEGATION\_KEYWORDS}}
	        \State \textbf{skip} $j$
	    \EndIf
	    \If{$j$ matches \textit{IGNORE\_SENT\_REGEX}}
	        \State \textbf{skip} $j$
	    \EndIf
	    \If{$j$ matches \textit{DEFINITE\_REGEX}}
	        \State \Return $\infty$
	    \EndIf
	    \If{$j$ matches \textit{POTENTIAL\_REGEX}}
		    \State $score = score + \textit{potential\_regex}(j) \times c$
	    \EndIf
	\EndFor
	\State $max\_score = max(max\_score, score)$
\EndFor
\State \Return $max\_score$
\end{algorithmic}
\end{algorithm}

\begin{figure*}
\fbox{\begin{minipage}{\textwidth}
\noindent
\texttt{"credit", "card", "scam", "fraud", "apple", "google", "samsung", "payment", "reward", "points"}. 
\end{minipage}}
\caption{The set of words we used to exclude certain web pages from the analysis. Such pages often mention tips for online protection, such as one-time codes sent to users for confirming payments.}
\label{fig:urlignore}
\end{figure*}

\begin{figure*}
\fbox{\begin{minipage}{\textwidth}
\texttt{r"(two|second|2)(.?)(factor|step)",}

\texttt{r"(otp|2fa|mfa|u2f)",}

\texttt{r"(trustfactor|duo|yubikey|yubico|fido|feitian|salt(.?)edge|rsa)"}
\end{minipage}}
\caption{URL keyword patterns to match.}
\label{fig:urlmatch}
\end{figure*}

\begin{figure*}
\fbox{\begin{minipage}{\textwidth}
\noindent
\texttt{"not", "none", "neither", "nowhere", "never", "without", "doesn't", "isn't", "wasn't", "shouldn't", "wouldn't", "couldn't", "won't", "can't", "don't", "does not", "is not", "was not", "should not", "would not", "could not", "cannot", "do not", "doesnt", "isnt", "wasnt", "shouldnt", "wouldnt", "couldnt", "wont", "cant", "dont"}.
\end{minipage}}
\caption{The set of negation words we used to identify if a banking site \textit{does not} support certain security features.}
\label{fig:negation}
\end{figure*}

\textbf{Scoring Algorithm.} The pseudo-code for calculating a 2FA-match score for a page is listed in Algorithm~\ref{alg:page.score}. Higher score means a higher likelihood that a site supports 2FA. 
For each bank website, we iterate over all the (URL-matched) pages with the goal of identifying the page with the maximum match score. Any page with its URL in the IGNORE\_LIST (Fig.~\ref{fig:urlignore}) is skipped as it is likely non-related (e.g., \texttt{"credit-card"} in the URL). Instead, if the URL of the page matches the criteria in URL\_REGEX (Fig.~\ref{fig:urlmatch}), we assign a score of 0.25 and continue. The next step involves pre-processing the raw HTML (using a Python parser~\cite{beautifulSoup}) to extract the text content and identify the individual sentences (using a sentence tokenizer in NLTK~\cite{bird2009natural}) in the page. In our analysis, we treat each sentence as a standalone input when searching for text patterns. To identify and skip sentences that explicitly mention that a security feature is not supported, we use the NEGATION\_KEYWORDS (Fig.~\ref{fig:negation}) list. We aim to reduce false positives (e.g., a site mentioning about one-time passcodes for credit card transactions) by skipping the sentence if it has one or more matches in IGNORE\_SENT\_REGEX (Fig.~\ref{fig:ignoresent}). We then search the sentence for unambiguous DEFINITE\_REGEX (Fig.~\ref{fig:defmatch}) matches that help us conclude a definite 2FA support (i.e., by returning a match score of $\infty$). For a page where we cannot find an unambiguous sentence on 2FA support, we count the number of matches (weighted by a constant) found in all the sentences matching the POTENTIAL\_REGEX (Fig.~\ref{fig:posmatch}) criteria. With a score for each page of the website, we find and return the page with the maximum match score for further analysis.

To identify areas where 2FA terms are commonly found within a bank site page, we manually reviewed 2FA pages on 50 bank sites. We selected from the banking sites listed in the crowd-sourced TwoFactorAuth project~\cite{twofactorauth} (discussed in Sec.~\ref{sec.related.work}) as it includes links (although unavailable for some sites) to individual site pages mentioning 2FA support. Based on our analysis, we assigned initial weights to various components of a web page and manually refined the scheme after reviewing scores for a subset of pages. 
In our analysis, we iteratively use the maximum-score matches to determine an empirical threshold above which the score suggests 2FA support, and also to evaluate accuracy.
It is also useful to consider the pattern matches for the lowest non-zero score as same minimum and maximum values might suggest higher confidence in the result (for both positive and negative matches). In this context, a false positive is when the search concludes that a site offers 2FA specifically for web logins when it does not. We utilize the false matches to improve our scoring scheme iteratively.

\begin{figure*}
\fbox{\begin{minipage}{1.025\textwidth}
\texttt{r"(credit|card|pay|payments?)",}

\texttt{r"\textsuperscript{$\wedge$}(?=.*(scams?|frauds?|fraudsters?))(?=.*?(ask|try|tries|share|steal))",}

\texttt{r"\textsuperscript{$\wedge$}(?=.*(your))(?=.*(username))",}

\texttt{r"\textsuperscript{$\wedge$}(?=.*(when|where))(?=.*(enabled|possible|available))",}

\texttt{r"\textsuperscript{$\wedge$}(?=.*(for))(?=.*(following|certain|critical|specific))(?=.*(transactions?|scenario))",}

\texttt{r"\textsuperscript{$\wedge$}(?=.*(we'll|will|future))(?=.*(implement|support|offer|deploy))"}
\end{minipage}}

\caption{Sentence patterns to ignore.}
\label{fig:ignoresent}
\end{figure*}

\begin{figure*}
\fbox{\begin{minipage}{\textwidth}
\texttt{r"\textsuperscript{$\wedge$}(?=.*(supports?|uses|offers?|deploy|implement))(?=.*?(2|two|multi)}

\hspace{6mm}\texttt{(.?)(factor|step)(.?)(auth|verification))",}

\texttt{r"\textsuperscript{$\wedge$}(?=.*(our|we|we'll|devices?))(?=.*?(uses?|supports?|offers?|generates?))}

\hspace{6mm}\texttt{(?=.*?(2fa|mfa|u2f|code|otp))",}

\texttt{r"\textsuperscript{$\wedge$}(?=.*(trustfactor|duo|yubikey|yubico|fido|feitian|salt(.?)edge|rsa))}

\hspace{6mm}\texttt{(?=.*?(push|app|auth|u2f|factor|token))"}
\end{minipage}}
\caption{Patterns for definite match.}
\label{fig:defmatch}
\end{figure*}

\begin{figure*}
\fbox{\begin{minipage}{\textwidth}
\texttt{r"(2|two|multi)(.?)(factor|step)(.?)(process|auth|verification)",}

\texttt{r"(one)(.?)(time|tap)(.?)(verification|code|password|passcode)",}

\texttt{r"(otp|2fa|mfa|u2f|token|authenticator)",}

\texttt{r"(trustfactor|duo|yubikey|yubico|fido|feitian|salt(.?)edge|rsa)",}

\texttt{r"\textsuperscript{$\wedge$}(?=.*(enter|input|authenticate))(?=.*(code|token|otp|))"}

\texttt{r"\textsuperscript{$\wedge$}(?=.*(activate|enhance|secure))(?=.*(account))",}

\texttt{r"\textsuperscript{$\wedge$}(?=.*(additional|enhanced?|extra))(?=.*(auth|protect|protection|security|check))"}
\end{minipage}}
\caption{Patterns for possible matches.}
\label{fig:posmatch}
\end{figure*}

\subsubsection{Patterns to exclude from matching}
\label{sec.exclude}
Before searching for positive matches, we use several lists of regular expressions to exclude certain patterns in both URLs (listed in Fig.~\ref{fig:urlignore}) and sentences (in Fig.~\ref{fig:ignoresent}). This order is important as it eliminates many potential false positive matches such as texts that describe one-time passwords for payment transactions (as opposed to web logins). As shown in Algorithm~\ref{alg:page.score}, we use a list of negation words (see Fig.~\ref{fig:negation}) to identify sentences that explicitly state lack of features.

\subsubsection{Patterns for score calculation}
For texts included for analysis, we use separate lists of regular expressions to calculate a match score for each sentence. If a sentence is a definite match (as defined in Fig.~\ref{fig:defmatch}), we conclude that the site supports 2FA. For more ambiguous patterns, we use the lists in Fig.~\ref{fig:posmatch} for calculating a score based on the number of related patterns found in the combination of all the text in the page.

\section{Preliminary Results}
\label{sec.empirical.study}
We now present our preliminary findings on 2FA support in online banking. We rely on the methodology discussed in the previous section to calculate match scores (for mentions of 2FA feature) per page of a bank domain. For each domain, we take the page with the maximum score for deducing whether 2FA is offered by the domain. We first look for phrases that definitively indicate that 2FA is offered. For example, if a sentence within a matched URL (e.g., URL contains the keyword ``security") includes the phrase ``we send a otp whenever you login", it is a definitive match. As discussed in Sec.~\ref{sec.exclude}, prior to searching for definite matches we exclude other contexts such as OTP for credit-card transactions. Our dataset includes 379 sites with one or more URL match. The findings on these sites discussed below is also listed in Table~\ref{tab:numberofsites}.

\begin{table}
\caption{Banking website support for 2FA. A manual check of the 15 of 68 sites gives an $\epsilon$ of 20\% for false positive estimate.}
\centering
\begin{tabular}{ll}
\toprule
Preliminary results & Number of banking websites \\
\midrule
Definite matches for 2FA & 68 - $\epsilon$ \\
No known matches found & 232 \\
Potential matches found & 79 \\            
\bottomrule
\end{tabular}
\label{tab:numberofsites}
\end{table}

\subsection{Definite matches}
We chose a random set of 10 sites that match our initial criteria to refine our search for definite matches. With the refined search, in total we found 68 sites that match our criteria for definite matches. In order to test the accuracy, we chose a random set of 15 sites among the definite matches (excluding the initial set of 10) and found 12 sites where the identified match was a true positive (i.e., site offers 2FA for online login). Of the three false positives (i.e., site does not offer 2FA for logins), two were instances where our exclusion criteria proved insufficient, e.g., page mentions use of OTP for password reset, but not for online login. Another instance involved incorrect parsing of text into sentences (i.e., the tokenizer library we used incorrectly treated one sentence as two).

\subsection{No match in the content}
Among the sites with one or more URL matches, we did not find a clear pattern in the content for 232 sites. We expect our dataset to include such sites as our criteria for URL matching includes strings that may appear in non-security pages. For example, ``2fa" is seen in many non-security URLs that contain encoded string such as filenames, UUIDs, and other identifiers. Another factor that could be influencing the high number is the total number of links our crawler visits for the domain, e.g., our crawler visits a relatively small number (of 100 links) for some domains compared to a typical bank domain (with 5,000 links). It is possible that our crawl optimizations (discussed in Sec.~\ref{sec.extracting.pages}) might be excluding the only links that lead to the site's security pages. However, we expect this to be minimal as most sites include any of their privacy, security, FAQ, and login pages in the site's index page, from which we begin our crawls. It may also be that the site is primarily in a different language with only a few English pages. Further analysis is required to identify other potential causes. 

\subsection{Potential matches}
We found 79 sites in our dataset with non-definitive match scores. For these sites, our search function did not find a clear matching sentence/phrase, but found patterns that might be related to 2FA. Using an initial constant value defined as $0.75*0.2$ for our scoring scheme (see Algorithm~\ref{alg:page.score}), we found 47 sites with a maximum score less than 1.0 and 32 sites with maximum scores greater than 1.0. A higher score means that multiple matching phrases were found across the page. It might also be useful to consider content across all relevant pages for a site. However, in this study we only consider each page as independent within the website, except when comparing the page scores to identify the maximum and minimum scores for the site.

We manually reviewed 15 sites with maximum scores of less than 1.0. We found only one instance\footnote{\url{https://www.absabank.mu/en/security-centre/}} where the page score is less than 1.0 whereas the page does state its 2FA feature. It was not captured as a initial definite match as this specific phrase was not included in our match criteria. Similarly, we selected a set of 10 sites with scores between 1.0 and 2.0, and 5 sites with scores greater than 2.0. All five sites with scores greater than 2.0 offer 2FA. We found two instances where a maximum page score in the range 1.0-1.40 when the sites do not offer 2FA. Pages of these sites provide incomplete information (for our analysis) in a single sentence. For example, a bank site\footnote{\url{https://mopss.mcb.com.pk/faq.jsp}} states that ``OTP has been introduced as an additional security feature", but suggests in a separate sentence that it is only offered for specific scenarios (e.g., for first login).

In another instance, we observed a (maximum) page score of 1.05 for a site page\footnote{\url{https://ally.com/do-it-right/trends/security-tips-for-your-personal-data/}} discussing security tips and also allowing users to add comments to the page. Our match patterns for this site identified a potential match for the user comment: ``\textit{It would be great if the bank can add Two-Factor Authentication functionality.}"

Our initial analysis suggests that when a definite match is not found for a site, a maximum page score of 1.50 or more indicates likely support for two-factor authentication. We also observe that there might be sites that use clear language stating support for 2FA, but our scheme finds a lower score (as discussed above). Identifying and including such additional patterns to our match criteria would improve our scheme.

\section{Related Work}
\label{sec.related.work}
While research literature on 2FA is extremely rich and covers numerous areas including usability, performance, and security, we are not specifically aware of research efforts towards automated enumeration of 2FA support in web authentication. It is not our intention to provide, herein, a comprehensive coverage of previous 2FA-related research, but we highlight relevant examples of such research. Ulqinaku et al.~\cite{ulqinaku2021is} discuss downgrade attacks in FIDO's U2F authentication scheme and show the possibility of in-the-wild attacks that use social engineering to exploit the vulnerability. Wang et al.~\cite{wang2016request} develop an evaluation criteria to measure the security properties offered by several 2FA schemes. Using the measurement framework, they compare 34 2FA proposals to identify the security requirements that each scheme satisfies. Wang et al.~\cite{wang2018measuring} use a similar framework to measure 2FA schemes used in industrial wireless sensor networks.

Related to usability, Krol et al.~\cite{krol2015they} present a study of user attitudes towards 2FA in online banking in the UK. Their findings identify usability issues involving the number of steps required when signing in with 2FA. To improve usability, the authors recommend the online banks to use similar wording for the same 2FA concepts (e.g., passcode vs. passphrase). More recently, a 2FA user study of online banking by Reese et al.~\cite{reese2019soups} finds participants who value security over inconvenience of 2FA for high value accounts such as online bank accounts. A related extension to our measurement study involving data on various usability factors found in online banks could offer additional insights regarding 2FA system designs.

The TwoFactorAuth~\cite{twofactorauth} project is a crowd-sourced effort providing information on sites with 2FA support. The dataset is built by members who contribute by submitting entries of specific sites and whether they support 2FA. While our goal is similar (i.e., identify whether sites offer 2FA), we use an automated methodology that is scalable to large number of sites. Petsas et al.~\cite{petsas2015two} measure user adoption of 2FA in over 100,000 Google accounts and find that only 6.4\% of accounts use the feature. Our endeavours herein can complement their work, as it can be extended to measure server-side support of 2FA over the web.

\section{Work in Progress}
\label{sec.future.work}
We now present our plans to advance this line of work.

\textbf{Improving our heuristics.} Our scoring algorithm for determining mentions of 2FA relies on heuristics we observed in a small subset of bank sites. Starting from an initial set, we refined our scoring scheme based on our findings. Our study also relies on excluding specific related patterns prior to considering other patterns in order to reduce false positives. In the next step, we plan to use a larger test sample to develop more robust heuristics to score the sentences within individual pages. We have already enumerated all banking websites from the Wikipedia (following our methodology in Sec.~\ref{sec:initiallist}), and found a total of 1,355 banks/websites. Additionally, we plan to explore the feasibility of extending our work to sites outside the banking domain (e.g., government sites).

\textbf{Using sentence structure analysis.} Our current methodology involves using pre-trained models to extract sentences from raw text. We then use a basic list of words to exclude sentences that contain negation words such as ``won't" and ``don't". Another approach might involve identifying the sentence structure (e.g., order of nouns and verbs in a sentence) for clearer interpretations of the meaning of sentences. Off-the-shelf models for sentence analysis~\cite[Ch.8]{bird2009natural} are insufficient as the problem of identifying 2FA phrases involves words that are interpreted as different parts of speech depending on the context. For example, the number word ``two" is often interpreted as an adjective, whereas in the context of 2FA it is part of the larger noun ``two-factor authentication".

\textbf{Topic modeling approach.} The problem of identifying sentence phrases that indicate support for 2FA is similar to processing raw text (e.g., user reviews in sentiment analysis) to extract the relevant interpretations. Chen et al.~\cite{chen2014ar} use a topic modeling framework to identify topics in user reviews in major mobile app marketplaces. For example, an informative review for the developer could be related to feature requests. In their framework, the authors pre-process raw reviews and feed the texts into a ranking model for computing scores for relevant topics in a given review. Our current study involves a similar approach of ranking sentences based on match scores for the 2FA features. A different approach might include a rule-based topic modeling to extend our scoring algorithm.

\section{Conclusion}
\label{sec.conclusion}
This study presents a work-in-progress towards a methodology for automated measurement of 2FA support by banking websites over the Internet. Our initial findings suggest that many online banks still lack support for 2FA. We hope that our empirical study on the banks that do offer 2FA encourages others to follow, and leads to a more secure financial ecosystem. Our goal involves only identifying whether sites offer 2FA. But a modified version of the selected text-analysis might help identify the type of 2FA (e.g., SMS OTP vs. hardware token).

\balance
\bibliographystyle{abbrv}
\bibliography{references}

\end{document}